\documentclass[twocolumn,aps,prl,superscriptaddress]{revtex4-1}
\usepackage{amssymb} 
\usepackage{amsmath,bm} 
\usepackage{graphicx} 
\usepackage[normalem]{ulem} 
\usepackage{multirow} 
\usepackage[colorlinks,linkcolor=blue,urlcolor=blue,citecolor=blue]{hyperref} 
\usepackage{lipsum} 
\usepackage[usenames, dvipsnames]{xcolor} 
\usepackage{tensor} 
\usepackage{isotope} 
\usepackage{amsmath} 

\usepackage{xspace}
\usepackage{epsfig} 
\usepackage{fancyhdr}
\usepackage{pslatex}   
\usepackage{xcolor}
\usepackage{color} 

\usepackage{lineno}

\allowdisplaybreaks[4] 

\newcommand {\hyt} {\mbox{$^3_\Lambda\text{H}$}\xspace}
\newcommand {\ahyt} {\mbox{${^3_{\bar{\Lambda}}}\overline{\rm H}$}\xspace}

\renewcommand{\sout}{\bgroup \color{red} \ULdepth=-.5ex \ULset} 
  
\begin{document} 
\title{Deciphering Hypertriton and Antihypertriton Spins from Their Global Polarizations in Heavy-Ion Collisions} 

\author{Kai-Jia Sun} 
\email{kjsun@fudan.edu.cn} 
\affiliation{Key Laboratory of Nuclear Physics and Ion-beam Application~(MOE), Institute of Modern Physics, Fudan University, Shanghai $200433$, China} 
\affiliation{Shanghai Research Center for Theoretical Nuclear Physics, NSFC and Fudan University, Shanghai 200438, China}

\author{Dai-Neng Liu} 
\affiliation{Key Laboratory of Nuclear Physics and Ion-beam Application~(MOE), Institute of Modern Physics, Fudan University, Shanghai $200433$, China} 
\affiliation{Shanghai Research Center for Theoretical Nuclear Physics, NSFC and Fudan University, Shanghai 200438, China}

\author{Yun-Peng Zheng} 
\affiliation{Key Laboratory of Nuclear Physics and Ion-beam Application~(MOE), Institute of Modern Physics, Fudan University, Shanghai $200433$, China} 
\affiliation{Shanghai Research Center for Theoretical Nuclear Physics, NSFC and Fudan University, Shanghai 200438, China}

\author{Jin-Hui Chen} 
\email{chenjinhui@fudan.edu.cn} 
\affiliation{Key Laboratory of Nuclear Physics and Ion-beam Application~(MOE), Institute of Modern Physics, Fudan University, Shanghai $200433$, China} 
\affiliation{Shanghai Research Center for Theoretical Nuclear Physics, NSFC and Fudan University, Shanghai 200438, China} 

\author{Che Ming Ko} 
\email{ko@comp.tamu.edu} 
\affiliation{Cyclotron Institute and Department of Physics and Astronomy, Texas A\&M University, College Station, Texas 77843, USA} 

\author{Yu-Gang Ma} 
\email{mayugang@fudan.edu.cn} 
\affiliation{Key Laboratory of Nuclear Physics and Ion-beam Application~(MOE), Institute of Modern Physics, Fudan University, Shanghai $200433$, China} 
\affiliation{Shanghai Research Center for Theoretical Nuclear Physics, NSFC and Fudan University, Shanghai 200438, China} 
 
\date{\today} 
\begin{abstract}  
Understanding the properties of hypernuclei is crucial for constraining the nature of hyperon-nucleon ($Y\text{-}N$) interactions, which plays a key role in determining the inner structure of compact stars. The lightest hypernuclei and antihypernuclei are the hypertriton ($^3_\Lambda\text{H}$), which consists  of a pair of nucleons and  a $\Lambda$ hyperon, and its antinucleus (${^3_{\bar{\Lambda}}}\overline{\rm H}$).  Significant knowledge has recently been acquired regarding the mass, lifetime, and binding energy of $^3_\Lambda\text{H}$. However, its exact spin, whether $\frac{1}{2}$ or  $\frac{3}{2}$, remains undetermined in both experimental and theoretical studies. Here, we present a novel method of using the hypertriton global polarization in heavy-ion collisions to decipher not only  its total spin but also its internal spin  structure.  This method is based on the finding that  its three different  spin structures  exhibit distinct beam energy dependence of its global polarization when it is produced in these collisions from the coalescence of proton, neutron and $\Lambda$. Future observations of the hypertriton and antihypertriton global polarizations thus provide the opportunity to unveil the spin structures of hypertriton and antihypertriton and their production mechanisms in heavy-ion collisions. 
\end{abstract} 

\maketitle 

\emph{Introduction}{\bf---}
Hypernuclei, bound states of hyperon(s) and nucleons,  are ideal probes to the hyperon-nucleon ($Y\text{-}N$) interaction, which is of fundamental importance for understanding the structure of compact astrophysical objects like the neutron stars and addressing the so-called ``hyperon puzzle"~\cite{Gal:2016boi,Saito:2021NRP, Ma_NST}. Understanding their properties is also important for validating  the fundamental theorem of charge, parity, and time reversal symmetry in quantum field theory~\cite{Chen:2018tnh,STAR:2011eej,ALICE:2015rey,STAR:2019wjm} and for the indirect search for dark matter in space~\cite{ALICE:2022zuz}. 

Although the hypernucleus was first discovered in 1952 using a nuclear emulsion cosmic
ray detector~\cite{Danysz1953}, the lightest antihypertriton (\ahyt)   was only detected in 2010 by the STAR Collaboration~\cite{STAR:2010gyg} at the Relativistic Heavy Ion Collider (RHIC).  Like \ahyt, the hypertriton (\hyt) is known to have a lifetime close to that of a free $\Lambda$ and a very small $\Lambda$ separation energy  $B_\Lambda=0.17 \pm 0.06$~MeV~\cite{Chen:2023mel,STAR:2021orx,ALICE:2022sco}, making its structure similar to that of a halo nucleus.  The spins of hypertriton and antihypertriton are, however, not precisely determined in  both  experimental    and  theoretical studies.  For hypertriton spin, it is usually determined by measuring the ratio of hypertriton's  partial decay widths $\Gamma_{^3\text{He}}$ into $^3{\rm He}+\pi^-$ and $\Gamma_{pd}$ into $p+d+\pi^-$, i.e., $R_3=\Gamma_{^3\text{He}}/(\Gamma_{^3\text{He}}+\Gamma_{pd})$~\cite{Dalitz:1959zz,Congleton:1992kk,Gal:2018bvq,Hildenbrand:2020kzu}. Although comparisons between model studies on $R_3$ and available experimental measurements favor the assignment $J^{\pi}=\frac{1}{2}^+$ over $J^{\pi}=\frac{3}{2}^+$,  the large errors in the experimental measurements  make this conclusion inconclusive~\cite{STAR:2017gxa}. Also, calculations based on the lattice QCD  with light-quark masses at the (unphysical) $\text{SU}(3)$-flavor symmetric point equal to the physical strange quark mass suggest the assignment of $J^{\pi}=\frac{3}{2}^+$ for the \hyt~\cite{NPLQCD:2012mex}.  

\begin{figure}[!t]
  \centering 
 \includegraphics[width=7cm]{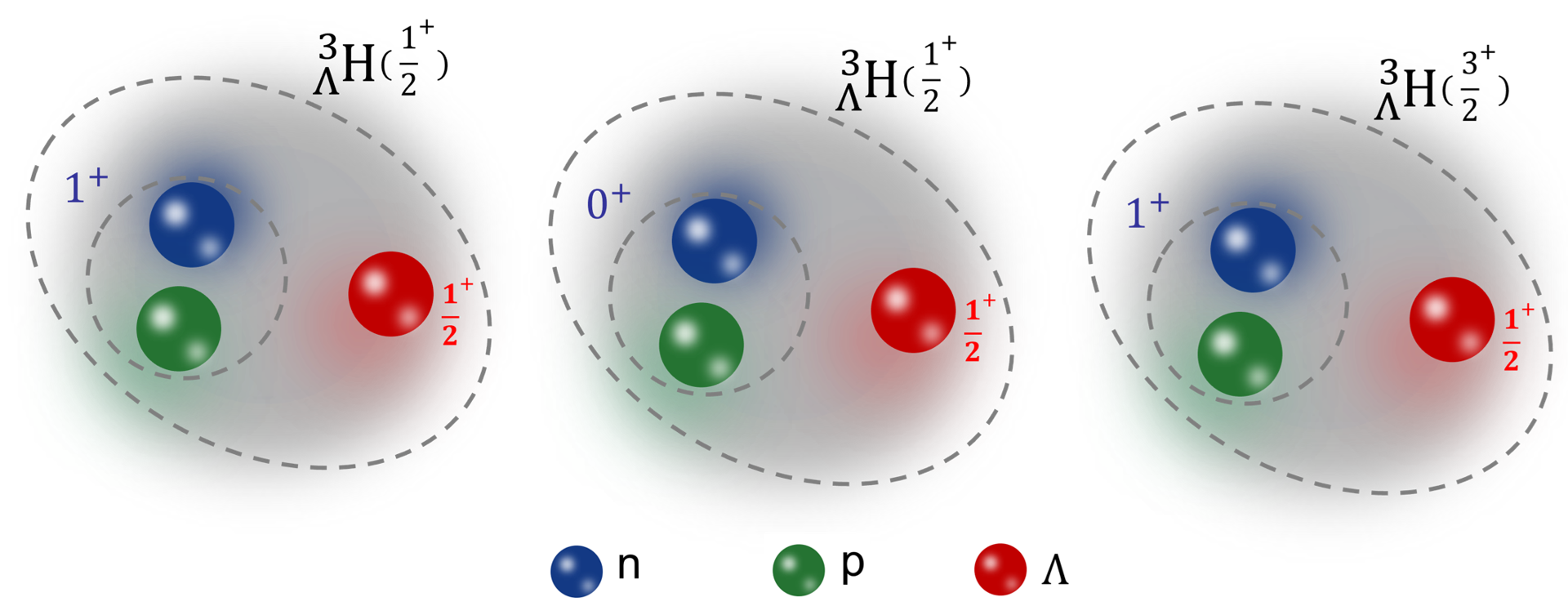}
  \caption{Hypothetical spin structure of hypertriton: spin-parity $J^{\pi}=\frac{1}{2}^+$ with the neutron-proton ($np$) pair in a spin-triplet state (left); $J^{\pi}=\frac{1}{2}^+$ with the $np$ pair in a spin-singlet state (middle); and $J^{\pi}=\frac{3}{2}^+$ with the $np$ pair in a spin-triplet state (right).}
  \label{pic:hyt}
\end{figure}

In the present study, we propose to  use measured global polarizations of hypertriton and antihypertriton in noncentral heavy-ion collisions to determine their spin structures. The study of hadron spin polarization in relativistic heavy-ion collisions has recently become an exciting and active field of research~\cite{Huang:2020xyr,Becattini:2020ngo,Chen:2023hnb,Becattini:2024uha,Niida:2024ntm}, particularly following the discovery of the spin polarization of hyperons ~\cite{STAR:2017ckg,STAR:2018gyt,STAR:2019erd,STAR:2020xbm} and the spin alignment of vector mesons~\cite{STAR:2022fan,ALICE:2019aid,ALICE:2022dyy}. Studies have shown that particles of nonzero spins can be globally polarized through their spin-orbit coupling in noncentral heavy-ion collisions where a very large orbital angular momentum is transferred to the produced quark-gluon plasma in the form of fluid vorticity along its direction~\cite{Liang:2004ph,Liang:2004xn,Voloshin:2004ha,Becattini:2007sr}. The values  of their polarizations are sensitive to the spin  structure of their constituent quarks according to the quark coalescence model for their production, in which the particle polarization is determined from the overlap of its spin-dependent Wigner function  with the spin-dependent quark distribution function in its emission source at hadronization~\cite{Liang:2004ph,Liang:2004xn}.  Similarly,  unstable light hypernuclei can acquire their polarizations  when they are produced from the coalescence of polarized nucleons and $\Lambda$ hyperons at the late stage of noncentral heavy-ion collisions~\cite{Liu:2023nkm}.  Like $\Lambda$ and $\bar{\Lambda}$~\cite{Lee:1957qs,Lee:1957he}, whose polarizations can be measured from their ``self-analyzing''  parity-violating weak decays~\cite{Dalitz:1958zz},  the polarizations of $^3_\Lambda\text{H}$ and $^3_{\bar{\Lambda}}\overline{\rm H}$ can be similarly measured from their weak decays. This makes it possible to utilize the hypertriton and antihypertriton global polarizations to decipher their total spins and internal spin  structures. 

To demonstrate the advantage of this new method, we consider three hypothetical spin structures of \hyt (see Fig.~\ref{pic:hyt}), including the spin-parity $J^{\pi}=\frac{1}{2}^+$ with the neutron-proton ($np$) pair in a spin-triplet state (left), $J^{\pi}=\frac{1}{2}^+$ with the $np$ pair in a spin-singlet state (middle), and $J^{\pi}=\frac{3}{2}^+$ with the $np$ pair in a spin-triplet state (right). We show that each  spin structure will result in a distinct beam energy dependence of \hyt and \ahyt polarizations in heavy-ion collisions from a few GeV to several TeV. 

\emph{Spin polarization of \hyt~ with spin-parity $J^{\pi}=\frac{1}{2}^+$}{\bf ---}
For the case that the \hyt is a loosely bound state of a spin half $\Lambda$ hyperon and a spin one deuteron ~\cite{Donigus:2020fon,Chen:2023mel}  as displayed in the left panel of Fig.~\ref{pic:hyt}, its spin wave function is given by
\begin{eqnarray}
\left|\frac{1}{2},\pm\frac{1}{2} \right\rangle_{\hyt} &=& \pm \frac{\sqrt{6}}{3} \left|\frac{1}{2},\pm\frac{1}{2}\right\rangle_{n}\left|\frac{1}{2},\pm\frac{1}{2}\right\rangle_{p}\left|\frac{1}{2},\mp\frac{1}{2}\right\rangle_\Lambda \notag \\
&\mp& \frac{\sqrt{6}}{6} \left(\left|\frac{1}{2},\frac{1}{2}\right\rangle_{n}\left|\frac{1}{2},-\frac{1}{2}\right\rangle_{p}\left|\frac{1}{2},\pm\frac{1}{2}\right\rangle_\Lambda\right.  \notag \\
&+&\left.\left|\frac{1}{2},-\frac{1}{2}\right\rangle_{n}\left|\frac{1}{2},\frac{1}{2}\right\rangle_{p}\left|\frac{1}{2},\pm\frac{1}{2}\right\rangle_\Lambda\right). 
\end{eqnarray}
The above equation shows that a hypertriton produced from the coalescence of   polarized proton, neutron, and $\Lambda$ hyperon can acquire a nonzero polarization. To see this quantitatively, we extend the blast-wave model of Ref.~\cite{Retiere:2003kf} to include the particle spin in 
its one-body phase-space distribution, i.e., 
\begin{eqnarray}
E_i\frac{\text{d}^3N_{i,\pm \frac{1}{2}}}{\text{d}{\bf p}_i^3} = \int_{\Sigma^\mu}\text{d}^3\sigma_\mu p_i^\mu w_{i,\pm\frac{1}{2}}({\bf x}_i,{\bf p}_i)\bar{f}_i({\bf x}_i,{\bf p}_i),
\end{eqnarray}
with $i$ denoting the particle species.  In the above equation, $\text{d}^3\sigma_\mu$ denotes the differential volume of the hypersurface $\Sigma^\mu$ of the emission source; $p_i^\mu$ denotes the particle four momentum; $w_{i,\pm \frac{1}{2}}= \frac{1}{2}[1\pm \mathcal{P}_i({\bf x}_i,{\bf p}_i)]$  with polarization $\mathcal{P}_i=\frac{N_{i,\frac{1}{2}}-N_{i,-\frac{1}{2}}}{N_{i,\frac{1}{2}}+N_{i,-\frac{1}{2}}}$  are the diagonal matrix elements of its spin density matrix, i.e., $\hat{\rho}_i={\rm diag}\left(\frac{1+\mathcal{P}_i}{2},\frac{1-\mathcal{P}_i}{2}\right)$; and $\bar{f}_i=\frac{g_i}{(2\pi)^3}\big{[} \exp(p_i^\mu u_\mu/T)/\xi_i+1 \big{]}^{-1}$ is the  spin-averaged phase-space   distribution function with $u^\mu$ being the local four flow velocity, and $g_i=2J_i+1$ and $\xi_i$ being the particle spin degeneracy and  fugacity factor, respectively.  In the absence of  correlations among  particle spins, the  spin density matrix of a $np\Lambda$ system  is given by $\hat{\rho}_{np\Lambda} = \hat{\rho}_n\otimes \hat{\rho}_p \otimes \hat{\rho}_\Lambda$. In this case, the Lorentz-invariant momentum distribution of hypertritons produced from the coalescence of polarized protons, neutrons, and lambdas can  be calculated from~\cite{Sun:2015ulc}
\begin{eqnarray}
\label{eq:Coal} 
E\frac{\text{d}^3N_{^3_\Lambda \text{H},\pm\frac{1}{2}}}{\text{d}{\bf P}^3} &=& E\int \prod_{i=n,p,\Lambda}p_i^\mu\text{d}^3\sigma_{\mu}\frac{\text{d}^3p_i}{E_i}\bar{f}_i({\bf x}_i,{\bf p}_i)  \notag \\
&\times&\left(\frac{2}{3}w_{n,\pm\frac{1}{2}}w_{p,\pm\frac{1}{2}}w_{\Lambda,\mp\frac{1}{2}}+\frac{1}{6}w_{n,\pm\frac{1}{2}}w_{p,\mp\frac{1}{2}}w_{\Lambda,\pm\frac{1}{2}}\right.\notag \\
&&+\left.\frac{1}{6}w_{n,\mp\frac{1}{2}}w_{p,\pm\frac{1}{2}}w_{\Lambda,\pm\frac{1}{2}}\right) \notag \\
&\times& W_{^3_\Lambda \text{H}}({\bf x}_n,{\bf x}_p,{\bf x}_\Lambda;{\bf p}_n,{\bf p}_p,{\bf p}_\Lambda) \delta({\bf P}-\sum_i{\bf p}_i),
\end{eqnarray}
where $W_{^3_\Lambda \text{H}}({\bf x}_n, {\bf x}_p, {\bf x}_\Lambda; {\bf p}_n, {\bf p}_p, {\bf p}_\Lambda)$ is the Wigner function of the \hyt internal wave function. The coalescence formula  based on the density matrix formulation~\cite{Scheibl:1998tk} has been shown to describe well \hyt production in high-energy nucleus collisions~\cite{Sun:2018mqq,Bellini:2018epz,ALargeIonColliderExperiment:2021puh}. Other approaches to light (hyper-)nuclei production include the statistical hadronization model~\cite{Andronic:2017pug}, the relativistic kinetic equations~\cite{Oliinychenko:2018ugs,Sun:2022xjr}, etc.

\begin{figure*}[!t]
  \centering 
 \includegraphics[width=18.0cm]{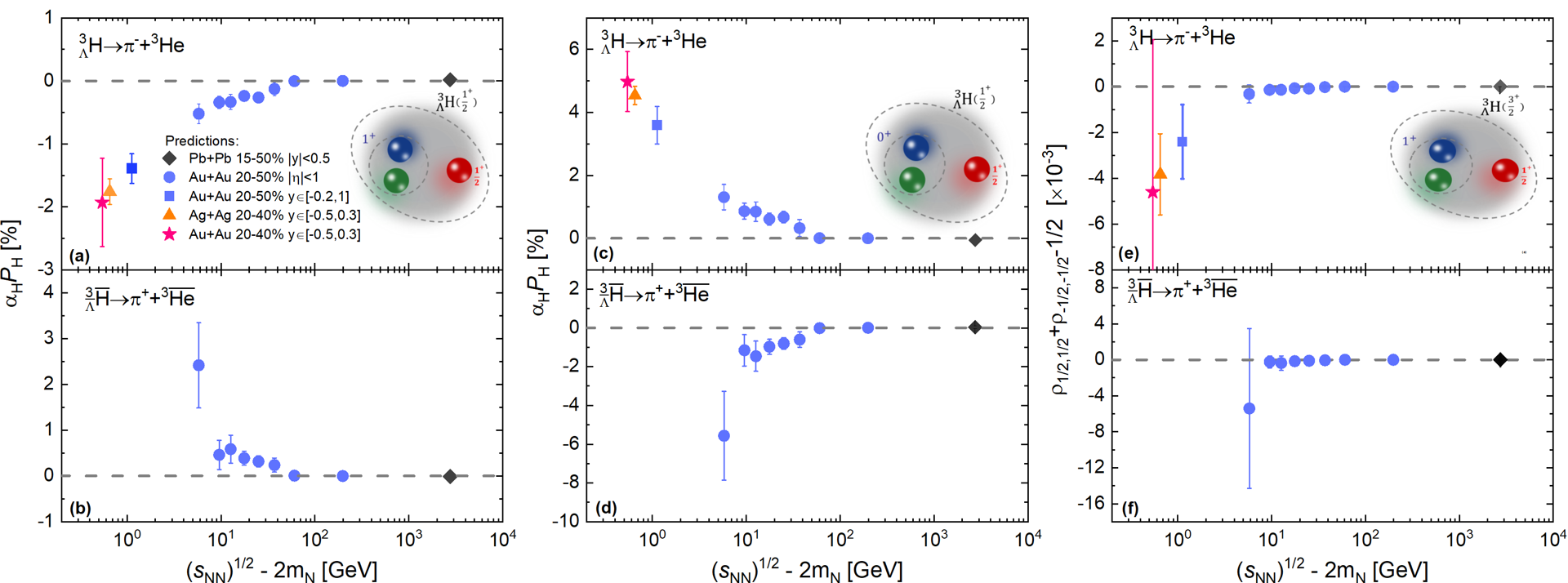}
  \caption{Collision energy dependence of hypertriton and antihypertriton polarizations predicted from  the decay channel  $^3_\Lambda\text{H}\rightarrow \pi^- +^3\text{He}$ (upper panels) and $\ahyt\rightarrow \pi^+ +^3\overline{\text{He}}$ (lower panels) based on the $\Lambda$ and $\bar{\Lambda}$ polarization data from the  STAR~\cite{STAR:2021beb}, ALICE~\cite{ALICE:2019onw}, and HADES~\cite{HADES:2022enx} Collaborations. Panels (a) and (b) depict results of $\alpha_H\mathcal{P}_H$ for \hyt~($\frac{1}{2}^+$) and \ahyt~($\frac{1}{2}^-$), respectively, assuming the  two nucleons inside the hypertriton and the two antinucleons inside the antihypertriton are in a spin-triplet state, while  results shown in panels (c) and (d)  are for the case that the two nucleons and two antinucleons are in a spin-singlet state.   Panels (e) and (f) depict results for the combined spin density matrix elements $\hat{\rho}_{\frac{1}{2},\frac{1}{2}}+\hat{\rho}_{-\frac{1}{2},-\frac{1}{2}}-\frac{1}{2}$ for \hyt ($\frac{3}{2}^+$) and \ahyt ($\frac{3}{2}^-$), respectively.}
  \label{pic:Spin}
\end{figure*}

For nucleon and $\Lambda$ hyperon polarizations that have small values and weak momentum dependence, which is the case from a study based on the chiral kinetic equations~\cite{Sun:2017xhx}, the hypertriton polarization obtained from Eq.~(\ref{eq:Coal}) is
\begin{eqnarray}
\label{eq:PolOneHalf}
\mathcal{P}_{^3_\Lambda \text{H}}&\approx& \frac{\frac{2}{3}\mathcal{P}_n+\frac{2}{3}\mathcal{P}_p-\frac{1}{3}\mathcal{P}_\Lambda-\mathcal{P}_n\mathcal{P}_p\mathcal{P}_\Lambda}{1-\frac{2}{3}(\mathcal{P}_n+\mathcal{P}_p)\mathcal{P}_\Lambda+\frac{1}{3}\mathcal{P}_n\mathcal{P}_p} \notag\\
&\approx& \frac{2}{3}\mathcal{P}_n+\frac{2}{3}\mathcal{P}_p-\frac{1}{3}\mathcal{P}_\Lambda.
\end{eqnarray}
Neglecting the small difference between the nucleon and $\Lambda$ polarizations due to the relativistic effect, the hypertriton then has the same global polarization as the $\Lambda$ hyperon, i.e., $\mathcal{P}_{^3_\Lambda \text{H}}\approx \mathcal{P}_\Lambda$. The same result  is obtained  if nucleons, $\Lambda$ hyperon, and \hyt~are in local thermal equilibrium as  the  polarization of a  nonrelativistic particle in this case is given by $\mathcal{P}=\frac{J+1}{3}\frac{w}{T}$ with $J$, $w$, and $T$ being the particle spin, vorticity, and local temperature, respectively~\cite{Becattini:2007sr,Becattini:2013fla,Becattini:2016gvu}. 

The global polarization of \hyt~  can be measured through its weak decays, such as $^3_\Lambda\text{H}\rightarrow \pi^-+^3\text{He}$, $^3_\Lambda\text{H}\rightarrow \pi^-+p+d$, $^3_\Lambda\text{H}\rightarrow \pi^-+n+p+p$,  etc.  We consider only the two-body decay channel $^3_\Lambda\text{H}\rightarrow \pi^-+^3\text{He}$ because it is easier to experimentally reconstruct the \hyt from its decay product. This decay can be viewed as a two-step process of $\Lambda$ decay ($\Lambda\rightarrow \pi^-+p$) followed by the recombination of the proton with the deuteron to form the $^3{\rm He}$, i.e., $p+d \rightarrow^3\text{He}$. The transition matrix for the process $\Lambda\rightarrow\pi^-+p$  is given by~\cite{Lee:1957he} 
\begin{equation}\label{angle}
 T(\Lambda\rightarrow \pi^-+p) = \frac{1}{\sqrt{4\pi}} 
 \left ( \begin{array}{cc}
 T_s+T_p\cos\theta_p^*& T_p\sin\theta_p^* e^{i\phi_p^*} \\
 T_p\sin\theta_p^* e^{-i\phi_p^*}&T_s-T_p\cos\theta_p^*
 \end{array} \right),
\end{equation}
where $T_s$ and $T_p$ are, respectively, the  $s$-wave and $p$-wave decay amplitudes for this process, and their values  have been determined to be $T_s \approx 0.917$ and $T_p \approx 0.399$ from the measured $\Lambda$ decay constant $\alpha_\Lambda = 2{\rm Re}(T_s^*T_p)=0.732\pm 0.014$~\cite{ParticleDataGroup:2022pth,BESIII:2022qax} if   their small imaginary parts are neglected.  For the decay constant of $\bar{\Lambda}$, we take $\alpha_{\bar{\Lambda}}=-\alpha_\Lambda$~\cite{BESIII:2022qax}. The  polar angles  $\theta_p^*$ and $\phi_p^*$ are   those between the momentum of the proton and the spin direction of $\Lambda$ hyperon in the $\Lambda$ rest frame.  If the recombination probability $F$ for the process $p+d\rightarrow^3$He, which depends on the overlap of the hypertriton wave function and the $\rm ^3He$ wave function, depends weakly on the proton angle in  the $\Lambda$ decay, the transition matrix for \hyt decay can then be approximately written as
\begin{eqnarray}
&&T(^3_\Lambda\text{H}\rightarrow\pi^-+^3\text{He})\notag\\
&& \hspace{1cm}=\frac{F}{6\sqrt{\pi}}\left (\begin{array}{cc}
 3T_s-T_p\cos\theta^*&  -T_p\sin\theta^*e^{i\phi^*}\\
 -T_p\sin\theta^*e^{-i\phi^*}&3T_s+T_p\cos\theta^*\end{array} \right),
\end{eqnarray}
where $\theta^*$ and $\phi^*$ are the polar angles of the $^3$He momentum with respect to the spin direction of \hyt, which are slightly different from the polar angles $\theta_p^*$ and $\phi_p^*$ in Eq.~(\ref{angle}). The normalized angular distribution of the $^3$He in the decay $^3_\Lambda\text{H}\rightarrow\pi^-+^3\text{He}$ is  given by
\begin{eqnarray}\label{triplet}
\frac{dN}{\sin\theta^* d\theta^*} &=&{\rm Tr}[T^+\hat{\rho} T] =  \frac{1}{2}(1+\alpha_{^3_\Lambda \text{H}}\mathcal{P}_{^3_\Lambda \text{H}}  \cos \theta^*)\notag\\
&\approx&  \frac{1}{2}\left(1-\frac{1}{2.58}\alpha_\Lambda\mathcal{P}_\Lambda \cos \theta^*\right),
\end{eqnarray}
in terms of the hypertriton decay parameter  $\alpha_{^3_\Lambda \text{H}}\approx-\frac{1}{3T_s^2+\frac{1}{3}T_p^2}\alpha_\Lambda \approx-\frac{1}{2.58}\alpha_\Lambda$.
Compared to the angular distribution of the proton in  $\Lambda$ decay, which has the form 
\begin{eqnarray}
\frac{dN}{\sin\theta_p^* d\theta_p^*} =\frac{1}{2}(1+\alpha_\Lambda\mathcal{P}_\Lambda \cos \theta_p^*), 
\end{eqnarray}
the $^3$He  in  \hyt decay has an opposite sign in its angular dependence. This  sign change in the angular distribution allows a precise determination of the internal spin structure of \hyt.

Using the measured polarizations of $\Lambda$ and $\bar{\Lambda}$ hyperon by the  STAR~\cite{STAR:2021beb}, ALICE~\cite{ALICE:2019onw}, and HADES~\cite{HADES:2022enx} Collaborations  allows us to predict  the energy dependence of $\alpha_H\mathcal{P}_H$ with $H$ denoting \hyt~or \ahyt through Eq.~(\ref{triplet}) for the case that the two nucleons inside the hypertriton and antihypertriton are in the spin-triplet state, and the results are shown in panels (a) and (b) of Fig.~\ref{pic:Spin}. It is seen that the magnitude of hypertriton and antihypertriton polarizations decreases with increasing collision energy, which is similar to the collision energy dependence of the  $\Lambda$ polarization measured in experiments~\cite{STAR:2017ckg} and calculated  in theoretical models~\cite{Huang:2020xyr,Becattini:2020ngo}. The magnitude of $\alpha_H\mathcal{P}_H$ reaches about 2\% in collisions at $\sqrt{s_{NN}}<7.7$ GeV where the hypertriton is more abundantly produced than in collisions at higher energies~\cite{Andronic:2010qu,STAR:2021orx}.

For the case that the $np$ pair inside the \hyt is in the spin-singlet state (see the middle panel of Fig.~\ref{pic:hyt}), the polarization of hypertriton is solely determined by that of the $\Lambda$ hyperon, i.e., $\alpha_{^3_\Lambda\text{H}}=\alpha_\Lambda$ and $\mathcal{P}_{^3_\Lambda\text{H}}=\mathcal{P}_\Lambda$.  The predicted energy dependence of $\alpha_H\mathcal{P}_H$ in this case are shown in panels (c) and (d) of Fig.~\ref{pic:Spin}, respectively, for \hyt and \ahyt, which are  very different from those shown in panels (a) and (b) of Fig.~\ref{pic:Spin}. These results clearly show that the spin polarizations of \hyt and \ahyt are sensitive to their internal spin structures.

\emph{Spin polarization of \hyt with spin-parity $J^{\pi}=\frac{3}{2}^+$}{\bf ---} For the case that \hyt~ has  the  spin-parity $J^P=\frac{3}{2}^+$ (see the right panel of Fig.~\ref{pic:hyt}),  its spin state can be described by a 4$\times$4 spin density matrix of unit trace. The diagonal elements of this matrix,  given by $\hat{\rho}_{\frac{3}{2},\frac{3}{2}}$, $\hat{\rho}_{\frac{1}{2},\frac{1}{2}}$, $\hat{\rho}_{-\frac{1}{2},-\frac{1}{2}}$, $\hat{\rho}_{-\frac{3}{2},-\frac{3}{2}}$, are probabilities for the spin component along a quantization axis to take the values of $3/2$, $1/2$, $-1/2$, and $-3/2$, respectively.   If a \hyt is produced from $n+p+\Lambda\rightarrow \hyt$ with $\mathcal{P}_n\approx\mathcal{P}_p\approx\mathcal{P}_\Lambda$,  we then have 
\begin{eqnarray}
&&\hat{\rho}_{^3_\Lambda\text{H}} \approx {\rm diag} \left[\frac{\left(1+\mathcal{P}_{\Lambda }\right){}^3}{4 \left(1+\mathcal{P}_{\Lambda }^2\right)},\frac{\left(1-\mathcal{P}_{\Lambda }\right) \left(1+\mathcal{P}_{\Lambda }\right){}^2}{4 \left(1+\mathcal{P}_{\Lambda }^2\right)},\right.\notag\\
&&\hspace{2cm}\left.\frac{\left(1-\mathcal{P}_{\Lambda}\right){}^2 \left(1+\mathcal{P}_{\Lambda }\right)}{4 \left(1+\mathcal{P}_{\Lambda }^2\right)},\frac{\left(1-\mathcal{P}_{\Lambda }\right){}^3}{4 \left(1+\mathcal{P}_{\Lambda }^2\right)}\right].
\end{eqnarray}
Following the method in Ref.~\cite{Becattini:2019ntv}, we obtain the   transition matrix for the two-body decay $^3_\Lambda\text{H}\rightarrow \pi^- +^3\text{He}$  as 
\begin{eqnarray}
\label{eq:TOneAndHalf}
&&T(^3_\Lambda\text{H}\rightarrow\pi^-+^3\text{He})\notag\\
&&\qquad= \frac{FT_p}{\sqrt{6\pi}} \left(
\begin{array}{cc}
 e^{i \phi^* } \sin\theta^* & 0 \\
 -\frac{2}{\sqrt{3}}  \cos\theta^*   & \frac{ e^{i \phi^* } \sin\theta^*   }{ \sqrt{3}} \\
 -\frac{ e^{-i \phi^* } \sin\theta^*   }{\sqrt{3}} & -\frac{2}{\sqrt{3}}  \cos\theta^*    \\
 0 & -e^{-i \phi^* } \sin\theta^*   \\
\end{array}
\right).
\end{eqnarray}
The normalized angular distribution  of $^3$He with respect  to  the \hyt spin direction is then given by
\begin{eqnarray}
\frac{dN}{\sin\theta^* d\theta^*} 
&=& \frac{1}{2} \left[1+  \left(\hat{\rho}_{\frac{1}{2},\frac{1}{2}}+\hat{\rho}_{-\frac{1}{2},-\frac{1}{2}}-\frac{1}{2}\right)(3\cos^2\theta^*-1 )\right], \notag \\
\end{eqnarray} 
with
\begin{eqnarray}
\hat{\rho}_{\frac{1}{2},\frac{1}{2}}+\hat{\rho}_{-\frac{1}{2},-\frac{1}{2}}-\frac{1}{2}\approx-\frac{\mathcal{P}_\Lambda^2}{1+\mathcal{P}_\Lambda^2}\approx -\mathcal{P}_\Lambda^2.
\end{eqnarray}
This angular distribution is analogous to the spin alignment of  a  vector meson~\cite{Chen:2023hnb,Becattini:2020ngo}.  The  predicted energy dependence of $\hat{\rho}_{\frac{1}{2},\frac{1}{2}}+\hat{\rho}_{-\frac{1}{2},-\frac{1}{2}}-\frac{1}{2}$ for \hyt and \ahyt using measured $\Lambda$ and $\bar{\Lambda}$ polarizations are depicted in panels (e) and (f) of Fig.~\ref{pic:Spin}, respectively. It is seen that the value of $\hat{\rho}_{\frac{1}{2},\frac{1}{2}}+\hat{\rho}_{-\frac{1}{2},-\frac{1}{2}}-\frac{1}{2}$ is close to zero for $\sqrt{s_{NN}}\ge 10$~GeV but reaches about $-4\times10^{-3}$ for $\sqrt{s_{NN}}-2m_N\le 1$ GeV.

The \hyt(\ahyt) spin-parity, structure, and decay mode as well as the angular distribution of the  $^3$He ($^3\overline{\text{He}}$) from the decay of \hyt(\ahyt) are summarized in Table~\ref{tab:spin}. 

\begin{table}[!t]
     \centering
     \begin{tabular}{|c|c|c|c|}
     \hline
         $J^{\pi}$& Structure &Decay mode & $dN/(\sin \theta^*d\theta^*)$   \\
        
         \hline
          $\frac{1}{2}^+$ & $\Lambda(\frac{1}{2}^+)-np(1^+)$ &$\hyt\rightarrow\pi^-+^3$He& $\frac{1}{2}(1-\frac{1}{2.58}\alpha_\Lambda\mathcal{P}_\Lambda\cos\theta^*)$  \\
          \hline
          $\frac{1}{2}^+$& $\Lambda(\frac{1}{2}^+)-np(0^+)$ &$\hyt\rightarrow\pi^-+^3$He&$\frac{1}{2}(1+\alpha_\Lambda\mathcal{P}_\Lambda\cos\theta^*)$    \\
          \hline
          $\frac{3}{2}^+$& $\Lambda(\frac{1}{2}^+)-np(1^+)$ &$\hyt\rightarrow\pi^-+^3$He&$ \frac{1}{2} \big{(}1-\mathcal{P}_\Lambda^2 (3\cos^2\theta^*-1)\big{)}$    \\ 
          \hline
          \hline 
          $\frac{1}{2}^-$ & 
          $\bar{\Lambda}(\frac{1}{2}^-)-\overline{np}(1^+)$ &
          $\ahyt\rightarrow\pi^++^3\overline{{\rm He}}$& 
          $\frac{1}{2}(1-\frac{1}{2.58}\alpha_{\bar{\Lambda}}\mathcal{P}_{\bar{\Lambda}}\cos\theta^*)$  \\
          \hline
          $\frac{1}{2}^-$& $\bar{\Lambda}(\frac{1}{2}^-)-\overline{np}(0^+)$ &$\ahyt\rightarrow\pi^++^3\overline{{\rm He}}$&$\frac{1}{2}(1+\alpha_{\bar{\Lambda}}\mathcal{P}_{\bar{\Lambda}}\cos\theta^*)$    \\
          \hline
          $\frac{3}{2}^-$& $\bar{\Lambda}(\frac{1}{2}^-)-\overline{np}(1^+)$ &$\ahyt\rightarrow\pi^++^3\overline{{\rm He}}$&$ \frac{1}{2} \big{(}1-\mathcal{P}_{\bar{\Lambda}}^2 (3\cos^2\theta^*-1)\big{)}$    \\
          \hline          
     \end{tabular}
     \caption{Hypertriton and antihypertriton spin-parity, structure, and decay mode as well as the angular distributions of the  $^3$He and $^3\overline{\text{He}}$ from their decays.}
     \label{tab:spin}
 \end{table}

\emph{Discussions}{\bf ---} 
The results shown in the above are based on the assumption that the hypertriton and antihypertriton have no excited states. If one or two spin states depicted in Fig. \ref{pic:hyt} are excited states of hypertriton, despite the lack of experimental evidence, or a spin $\frac{1}{2}$ hypertriton is a  mixture of the two spin $\frac{1}{2}$ structures in Fig. \ref{pic:hyt}, simple estimates based on the formalism for calculating the feed-down contribution to $\Lambda$ polarization in Ref.~\cite{Becattini:2016gvu} indicate that, although the feed-down contribution from the electromagnetic decay of the excited states of hypertriton can affect the magnitude of measured hypertriton polarization, its collision energy dependence in heavy-ion collisions remains qualitatively similar to that shown in Fig. \ref{pic:Spin}.

Also, the above results are obtained without including possible correlations among the spins of nucleons and the $\Lambda$ hyperon.  For  the spin alignment of $\phi$ meson, it has been shown in Refs.~\cite{Sheng:2022wsy,Lv:2024uev} that a strong negative  correlation between the spins of  $s$ and $\bar{s}$ quarks can lead to a significant enhancement of its value. To see   the effect of spin-spin correlations on the hypertriton polarization, we consider the case that its  spin structure is given by the one in the left panel of Fig.~\ref{pic:hyt}, i.e., $J^{\pi}=\frac{1}{2}^+$ with the $np$ pair in a spin-singlet state. Following the study of the effects of density fluctuations and  correlations on light nuclei production~\cite{Sun:2017xrx,Sun:2020zxy}, the density matrix of $np\Lambda$ can be generally expanded as~\cite{Lv:2024uev,Sheng:2020ghv}
\begin{eqnarray}
&&\hat{\rho}_{np\Lambda} = \hat{\rho}_n\otimes \hat{\rho}_p \otimes \hat{\rho}_\Lambda+\frac{1}{2^2}(c_{np}^{\alpha\beta}\hat{\sigma}_{n,\alpha}\otimes \hat{\sigma}_{p,\beta}\otimes \hat{\rho}_{\Lambda}\notag\\
&&\hspace{1cm}+c_{p\Lambda}^{\alpha\beta}\hat{\sigma}_{p,\alpha}\otimes \hat{\sigma}_{\Lambda,\beta}\otimes \hat{\rho}_{n} +c_{n\Lambda}^{\alpha\beta}\hat{\sigma}_{n,\alpha}\otimes \hat{\sigma}_{\Lambda,\beta}\otimes \hat{\rho}_{p})\notag\\
&&\hspace{1cm}+\frac{1}{2^3}c_{np\Lambda}^{\alpha\beta\gamma}\hat{\sigma}_{n,\alpha}\otimes\hat{\sigma}_{p,\beta}\otimes\hat{\sigma}_{\Lambda,\gamma},
\end{eqnarray}
where $\hat{\sigma}$ are the Pauli spin matrices, and $c_{ij}^{\alpha\beta}$ and $c_{ijk}^{\alpha\beta\gamma}$ are the spin correlation coefficients~\cite{Lv:2024uev}. The  hypertriton polarization given in Eq.~(\ref{eq:PolOneHalf}) is then modified to 
\begin{eqnarray}
\label{eq:pola}
 \mathcal{P}_{^3_\Lambda \text{H}}&\approx& \frac{\frac{2}{3}\langle\mathcal{P}_n\rangle+\frac{2}{3}\langle\mathcal{P}_p\rangle-\frac{1}{3}\langle\mathcal{P}_\Lambda\rangle-\langle\mathcal{P}_n \mathcal{P}_p\mathcal{P}_\Lambda\rangle+C_-}{1-\frac{2}{3}(\langle(\mathcal{P}_n+\mathcal{P}_p)\mathcal{P}_\Lambda\rangle)+\frac{1}{3}\langle\mathcal{P}_n\mathcal{P}_p\rangle+C_+},
\end{eqnarray}
where the  two ``genuine" correlation terms are given by 
\begin{eqnarray}
\label{eq:polb}
C_- &=& -\frac{1}{4}(\langle c_{np}^{zz}\mathcal{P}_\Lambda \rangle+\langle c_{p\Lambda}^{zz}\mathcal{P}_n \rangle+ \langle c_{n\Lambda}^{zz}\mathcal{P}_p\rangle) -\frac{1}{4}\langle c_{np\Lambda}^{zzz}\rangle,   \notag  \\
C_+ &=& \frac{1}{12}(\langle c_{np}^{zz}\rangle-2\langle c_{p\Lambda}^{zz}\rangle-2\langle c_{n\Lambda}^{zz}\rangle).     
\end{eqnarray}
The $\langle \cdots \rangle$ in the above equations denotes the average over the nucleon and $\Lambda$ hyperon spin-dependent phase-pace distributions weighted by the spin-dependent hypertriton Wigner function as in Eq.~(\ref{eq:Coal}).     Equations (\ref{eq:pola})  and (\ref{eq:polb}) show that the polarization of \hyt depends on both the two-body and the three-body spin correlation. A negative or positive spin correlation among $\Lambda$ hyperon and nucleons would lead to an enhanced or suppressed hypertriton polarization.  The effect is, however, not large  if it is estimated from Eq.~(\ref{eq:pola}) by taking the two-body correlation terms to be identical, neglecting the three-body correlation terms and terms like $\langle\mathcal{P}_i\mathcal{P}_j\rangle$, and letting all $c^{zz}$ to have values similar to   the largest $\Lambda$ and $\bar\Lambda$ spin correlation extracted in Ref.~\cite{Lv:2024uev} from the $\phi$ meson spin alignment measured in STAR experiments.

\emph{Summary}{\bf ---}
In the present study, we propose  to decipher the spin structures of hypertriton and antihypertriton by measuring their global polarizations in noncentral heavy-ion collisions.
This new approach enables a precise determination of both the total spin and the internal spin structures of hypertriton and antihypertriton.

We first demonstrate that hypertriton and antihypertriton can acquire global polarizations through the coalescence of polarized nucleons and $\Lambda$ hyperons at the kinetic freeze-out of these collisions. We then show that both the magnitude of hypertriton polarization and the angular distribution of the product from its two-body decay $^3_\Lambda\text{H}\rightarrow \pi^-+^3\text{He}$ strongly depend on its total spin and internal spin structure. For $^3_\Lambda\text{H}$ with spin-parity $J^{\pi}=\frac{1}{2}^+$, the angular distribution of  $^3$He from its two-body decay has a $\cos \theta^*$ dependence. There is, however, a sign change in the angular distribution for the $np$ pair in a spin-triplet state or spin-singlet state.  For a $^3_\Lambda\text{H}$ with  $J^{\pi}=\frac{3}{2}^+$,  the angular distribution of $^3$He from its decay has, on the other hand, a $\cos 2\theta^*$ dependence, and its combined spin density matrix elements $\hat{\rho}_{\frac{1}{2},\frac{1}{2}}+\hat{\rho}_{-\frac{1}{2},-\frac{1}{2}}-\frac{1}{2}$ has a small negative value. 

For each hypertriton or antihypertriton spin structure shown in Fig.~\ref{pic:hyt},  a distinct beam energy dependence of their polarizations is predicted for heavy-ion collisions from a few GeV to several TeV.  These results can be readily confronted with upcoming experimental measurements from HADES, STAR, ALICE, and other collaborations, and the comparisons will allow us to decipher the spin structures of hypertriton and antihypertriton and give insights to the nature of $Y\text{-}N$ interactions. 

\begin{acknowledgments}  
This work was supported in part by the Natural Science Foundation of Shanghai under Grant No.\ 23JC1400200, the National Key Research and Development Project of China under Grant No. 2024YFA1612500, No.\ 2022YFA1602303, the National Natural Science Foundation of China under Contracts No.\ 12422509, No.\ 12375121, No.\ 12025501, No.\ 12147101, No.\ 11891070, No.\ 124B2102, the Guangdong Major Project of Basic and Applied Basic Research No.\ 2020B0301030008, the STCSM under Grant No.\ 23590780100, and the U.S.\ Department of Energy under Award No.\ DE-SC0015266. 
\end{acknowledgments}  

%

\end{document}